\documentclass[%
 reprint,
 amsmath,amssymb,
 aps,
]{revtex4-1}
\usepackage{afterpage}
\usepackage{color}
\usepackage{graphicx}
\usepackage{dcolumn}
\usepackage{bm}
\usepackage{slashed}

\newcommand{\comment}[1]{}

\begin{document}

\preprint{APS/123-QED}

\title{Protection against Spin Gap in 2-d Insulating Antiferromagnets with a Chern-Simons Term}

\author{Imam Makhfudz and Pierre Pujol}
\affiliation{%
Laboratoire de Physique Th\'{e}orique--IRSAMC, CNRS and Universit\'{e} de Toulouse, UPS, F-31062 Toulouse, France 
}%
\date{\today}

\begin{abstract}

We propose a novel mechanism for the protection against spin gapped states in doped antiferromagnets. It requires the presence of a
Chern-Simons term that can be generated by a coupling between spin and an insulator.
We first demonstrate that in the presence of this term the vortex loop excitations of the spin sector behave as anyons with fractional statistics.
To generate such term, the fermions should have massive Dirac spectrum coupled to the emergent spin field of the spin sector.
The Dirac spectrum can be realized by a planar spin configuration arising as the lowest-energy configuration of a square lattice antiferromagnet Hamiltonian 
involving a Dzyaloshinskii-Moriya interaction. The mass is provided by a combination of
dimerization and staggered chemical potential. We finally show that for realistic parameters, anyonic vortex loop condensation will likely never occur and
thus the spin gapped state is prevented. We also propose real magnetic materials for an experimental verification of our theory.




\end{abstract}

\pacs{Valid PACS appear here}
\maketitle


\section{Introduction}
The discovery of high $T_c$ superconductivity in Cuprates led to a flurry of new ideas such as the idea of spin liquid state as one of the possible explanations
for the emergence of superconductivity from Mott insulating parent compounds \cite{AndersonRVB} and the importance of the physics of doped antiferromagnets \cite{LeeNagaosaWen}. 
An exotic spin liquid state from Heisenberg types of spin models involving breaking 
of discrete symmetries such as parity or time-reversal symmetry, the so-called chiral spin liquid, has been proposed \cite{KalmeyerLaughlin}. The effective low energy theory of such chiral spin state
normally involves a topological term called Chern-Simons term \cite{W-W-Z}. It was originally studied in particle physics \cite{DeserJackiwTempleton} and mimicks the fractional quantum Hall effect (FQHE) \cite{Laughlin} where such term 
appears as the low energy effective theory in the bulk \cite{ZHK}. 

The Chern-Simons term can be generated by a fermion-gauge field coupling when the fermion is integrated out. One obtains a fermion determinant which gives the Chern-Simons term as the action for gauge field $A$ 
\cite{Schaposnik}

\begin{equation}\label{CSfromfermion}
 S_{CS}[A]=i\frac{N_f}{2}\frac{e^2}{4\pi}\frac{m_{\psi}}{|m_{\psi}|}\int d^3x\epsilon^{\mu\nu\lambda}A_{\mu}\partial_{\nu}A_{\lambda}
\end{equation}
in Euclidean space-time upon perturbative expansion \cite{Caution}, where $N_f$ is the number of fermion flavors, $e$ is the gauge charge, $m_{\psi}$ is the Dirac fermion mass, and $\epsilon^{\mu\nu\lambda}$ is totally antisymmetric tensor. 
The Chern-Simons action depends only on the overall sign of the mass rather than its magnitude even though the mass must be nonzero for the expansion to make sense.

In quantum magnetism, the magnetization curve may show the presence of plateaus.
It is well understood that the plateau state corresponds to a state with gapped magnetic excitations while outside the plateau, the spin sector is gapless \cite{TTH}. 
In doped antiferromagnets \cite{LCP}, this leads to distinct natures of interaction between the fermions \cite{plateauIM-PP1}. 
A state with gapped magnetic excitations is normally associated with preserved continuous symmetry with no long range magnetic order in the direction transverse to the applied magnetic field \cite{TTH}.
In this work, we show that in the presence of a Chern-Simons term one can have either one of two scenarios: i)A spin gapped state occurs as the analog of the fractional quantum Hall effect (FQHE) with its chiral edge states
ii)The system is protected against such spin gapped state. In this paper, we consider 2-d antiferromagnets on square lattice, as we did in our previous work \cite{plateauIM-PP1}. 

In this paper, we also propose a way to realize a chiral theory as an effective low-energy theory of spin systems explicitly rather than spontaneously by considering doped antiferromagnets on the square lattice.
The fermions hop on the lattice on top of a pre-existing magnetic background. Because of a strong Hund coupling, the spin of the electron must be parallel to that of the local spin within the adiabatic approximation.
The resulting dynamics is well-described by effective tight-binding Hamiltonian \cite{Ohgushi}
\begin{equation}\label{tightbindingH}
 H=-t\sum_{\mathbf{r},\mathbf{r}'}\langle \Omega_{\mathbf{r}}|\Omega_{\mathbf{r}'}\rangle c^{\dag}_{\mathbf{r}'}c_{\mathbf{r}}+h.c.
\end{equation}
where the spin sector modifies the hopping integral via the overlap of spin coherent states $|\Omega_{\mathbf{r}}\rangle$ between nearest-neighbor sites, a mechanism well established 
from the studies of anomalous Hall effect \cite{AHErmp} and doped antiferromagnets \cite{ShankarPRL}. 
This coherent state overlap has two effects: 
providing the background flux for the fermions and the effective spin field that couples to the fermions. The spin sector should take an appropriate classical 
lowest-energy configuration which produces a staggered $\pi$-flux state \cite{AffMarston+Kotliar} with dimerization and staggered chemical potential 
needed to get a massive Dirac fermion spectrum.

\section{Field Theory}
We use the semiclassical spin path integral approach \cite{TTH} and start with the Euclidean effective action of 2-d doped antiferromagnet 

\begin{equation}\label{spinsectoraction1}
 S_{\phi}=\int d^2 x \int d\tau \frac{K_{\tau}}{2}(\partial_{\tau}\phi)^2 + \frac{K_{r}}{2}(\nabla\phi)^2 + i\left(\frac{S}{a^2}\right)\partial_{\tau}\phi
\end{equation}
\begin{equation}\label{holeaction}
 S_{\overline{\psi},\psi}=\int d^2 x \int d\tau\overline{\psi}[\gamma^{\mu}(-i\partial_{\mu}-eA_{\mu})+m_{\psi}]\psi
\end{equation}
taking the form of an $XY$ model describing quantum fluctuations around a classical lowest-energy configuration specified by $\mathbf{S}=S(\sin\theta_0\cos\phi_0,\sin\theta_0\sin\phi_0,\cos\theta_0)$ 
plus a Berry phase action whose form follows from the spin configuration we will describe. 
The $\phi$ is the quantum fluctuating phase angle field around $\phi_0$ \cite{TTH} and is conjugate to the momentum operator $\Pi$; 
$[\phi_{\mathbf{r}},\Pi_{\mathbf{r}'}]=i\delta_{\mathbf{r}\mathbf{r}'}$ where $\Pi$ can be defined in terms of $\theta$ \cite{LCP}.
The $K_{\tau},K_{r}$ are stiffness coefficients of the spin sector.
The $\overline{\psi},\psi$ are Dirac spinors representing electrons that couple to the spin sector's $\phi$ field via the pseudo-gauge field given as $A_{\mu}=\partial_{\mu}\phi,\mu=\tau,x,y$. 
Here two comments are in order. First, the definition of the vector field $A_{\mu}$ looks as if it were a pure gauge. 
However, one has to keep in mind that the field $\phi$ may have vorticity, which turns out to be at the origin of a non-zero flux for  $A_{\mu}$. 
Second, the action \label{spinsectoraction} is not gauge invariant and as such the theory is not a genuine gauge theory. 
We use nevertheless the terminology of pseudo-gauge field for $A_{\mu}$, which we also refer to as spin field, because of the way it couples to the charge degrees of freedom.
When the fermions are integrated out, we get the Chern-Simons term due to the fermion-spin field coupling given in Eq. (\ref{CSfromfermion}). 
A similar final action but without Berry phase has been studied in a different context \cite{grapheneDiracmass}. 

Integrating out the massive Dirac fermions in Eq.(\ref{holeaction}), we obtain a Chern-Simons term in terms of the phase field $\phi$
\begin{equation}\label{CS}
 \mathcal{L}_{CS}=i\frac{\kappa}{2\pi}\epsilon^{\mu\nu\lambda}\partial_{\mu}\phi\partial_{\nu}\partial_{\lambda}\phi
\end{equation}
with $\kappa=e^2/4$ ($N_f=1$). 
We first investigate the effect of this Chern-Simons term added to the action Eq.(\ref{spinsectoraction1}) by applying a duality transformation which introduces a Hubbard-Stratonovich 
auxiliary vector field $J_{\mu}$ \cite{TTH} and re-expresses the full action as 

\begin{equation}\label{XYBPCS}
 \mathcal{L}=\frac{J^{\mu}J_{\mu}}{2K_{\mu}}+i(J^{\mu}+\frac{S}{a^2}\delta^{\mu}_{\tau})\partial_{\mu}\phi+i\kappa \partial_{\mu}\phi J^{\mu}_{\mathcal{V}}
\end{equation}
In this case, we have decomposed the phase field into vortexful and regular parts $\phi=\phi_{\mathcal{V}}+\phi_R$ and defined vortex loop current 
$J^{\lambda}_{\mathcal{V}}=(1/(2\pi))\epsilon^{\lambda\mu\nu}\partial_{\mu}\partial_{\nu}\phi_{\mathcal{V}}$. 
Following boson-vortex duality transformation \cite{TTH} and working in Euclidean space-time, we obtain the following effective Lagrangian for $J^{\mu}_{\mathcal{V}}$
\[
L[J_{\mathcal{V}}]= \int_{k} J^{\mu}_{\mathcal{V}}(k)\frac{1}{k^2}\left(\left(\delta_{\mu\nu}-\frac{k_{\mu}k_{\nu}}{k^2}\right)-\kappa \pi \epsilon_{\mu\nu\alpha}k^{\alpha}\right)J^{\nu}_{\mathcal{V}}(-k)
\]
\begin{equation}\label{effeactionJv}
+i2\pi\int_x\overline{b}_{\mu}(x)J^{\mu}_{\mathcal{V}}(x)
\end{equation}
where $\overline{b}_{\mu}=\frac{1}{2}(\frac{S}{a^2})\epsilon_{\tau\mu\nu}x_{\nu}$. 
As in \cite{TTH}, the above result was obtained under the assumption that only configurations with closed 
vortex loops do contribute to the low energy physics. 
The effective magnetic field through a vortex loop is given by

\begin{equation}\label{newgaugefieldflux}
 \epsilon^{\mu\nu\lambda}\partial_{\nu}\hat{b}_{\lambda}=\frac{S}{a^2}\delta^{\mu}_{\tau}+\kappa J^{\mu}_{\mathcal{V}}
\end{equation}
We note that we get an extra term $\kappa J^{\mu}_{\mathcal{V}}$ to the effective magnetic field coming from the Chern-Simons term.
The important consequence of this is that the resulting effective Berry phase for a vortex loop now has a total contribution coming from ordinary Berry phase term and Chern-Simons term and 
is given by

\begin{equation}\label{vortexflux}
 \Phi_{\mathrm{vortex\, loop}}=2\pi S\frac{A}{a^2} q+2\pi q \kappa \Phi_{\mathrm{other\, vortex \, loop}}
\end{equation}
where $A$ is the area of a vortex loop and equals integer multiple of $a^2$, $q$ is the vorticity of the vortex loop, and $\Phi_{\mathrm{other\, vortex\,loop}}$
is the flux of other vortex loop, nonzero only if that other vortex loop is knotted across the first vortex loop.
Comparing Eqs. (\ref{newgaugefieldflux}) and (\ref{vortexflux}) suggests that the Chern-Simons contribution is nonzero
only if the two vortex loops are linked.

We verify the above proposition as follows.
For two nonintersecting curves $\gamma_1,\gamma_2$ (which act as mapping from manifold $\mathbb{S}_1$ (circle) to 3-d Cartesian space $\mathbb{R}^3$ (the 3-d Euclidean space in our problem), 
the linking number is given by

\begin{equation}\label{linking-number}
 \mathcal{N}_{\mathrm{linking}}=\frac{1}{4\pi}\oint_{\gamma_1}\oint_{\gamma_2}\frac{(\mathbf{r}_1-\mathbf{r}_2)\cdot(d\mathbf{r}_1\times d\mathbf{r}_2)}{|\mathbf{r}_1-\mathbf{r}_2|^3}
\end{equation}
Taking the inverse Fourier transform, with the vortex current given by $\mathbf{J}_{\mathcal{V}}=q d\mathbf{r}$ and endowing the above expression with an overall coefficient $\kappa$, 
the above expression corresponds to vortex current-vortex current interaction 

\begin{equation}\label{dualcurrentnetaction}
 L_{\mathrm{linking}}[J_{\mathcal{V}}]=\int_k J^{\mu}_{\mathcal{V}}(k)\frac{1}{k^2}\left(-\kappa \pi \epsilon_{\mu\nu\alpha}k^{\alpha}\right)J^{\nu}_{\mathcal{V}}(-k)
\end{equation}
in Fourier space. This is precisely equal to the topological part of Eq.(\ref{effeactionJv}). This implies that the Chern-Simons term leads to the linking between vortex-loops, resulting in nonzero linking number, precisely as implied by 
Eqs. (\ref{newgaugefieldflux}) and (\ref{vortexflux}). 

\section{Anyon Physics}
We now show how the field theory and the results in the previous section lead to the anyonic fractional statistics of the vortex loops excitations of the 2-d $XY$ model Eq.(\ref{spinsectoraction1}).
Without the Chern-Simons term, the $XY$ model Eq.(\ref{spinsectoraction1}) has topological excitations in the form of vortex loops
which, in the absence of destructive interference due to the Berry phase, can proliferate. The vortex loops can be considered as worldlines of Bose particles and the proliferation of vortex loops 
corresponds to a condensation of the bosons. This vortex proliferation implies disordering of the dual ($\phi$) field, which gives rise to gapped magnetic excitations.  
In the presence of the Chern-Simons term, vortex loops become anyons and a number of vortex loops must collect together in order to form a boson, and then be able to condense. 
One then can have either one of the two scenarios: i)The anyonic
vortex loops are able to form bosons and condense, giving rise to a state with gapped magnetic excitations as the analog of FQHE state, 
ii) The anyons fail to condense, for which a state with gapped magnetic excitations will not occur. 

To show this technically in a simple picture, consider a system of two vortex loops with vorticities $q_1$ and $q_2$ respectively knotted to each other once ($\mathcal{N}_{\mathrm{linking}}=1$).
The partition function of this system has a contribution of the form 

\begin{equation}\label{partitionfunction}
 e^{i 4\pi\kappa q_1q_2 +  q_1 q_2 E_{\mathrm{Coulomb}}}
\end{equation}
Apart from the Coulomb interaction, the contribution to the partition function given in Eq. (\ref{partitionfunction}) resembles that of linked wordlines of anyon system with its fractional statistics \cite{fractionalstat}. 
The first term in the exponent $4\pi\kappa q_1 q_2$ equals twice the statistical angle $\Theta$ under the exchange of two anyons realized by vortex loops with vorticities $q_1$ and $q_2$, 
giving $\Theta=2\pi\kappa q_1 q_2$. In 3-d Euclidean space-time, there exists non-trivial braiding statistics between loops \cite{Preskill}. Our system thus manifests anyonic loop statistics in 3-d.
With $q_1=q_2=1$ for elementary vortex loop, we obtain $\Theta=2\pi\kappa$ and therefore need $\mathcal{N}_{\mathrm{anyon}}=2\pi/\Theta=1/\kappa$ anyons to get a boson before the anyons can condense. 
The occurrence of anyon condensation (and that of spin gapped state) thus depends crucially on the magnitude of the Chern-Simons coupling $\kappa$. 
The Coulomb interaction then determines the critical value of the parameter that characterizes any possible condensation transition for this interacting system.
We will show a scenario from a microscopic model where $\kappa$ takes small enough values, leading to large $\mathcal{N}_{\mathrm{anyon}}$ and as a consequence prevents anyon condensation 
and the occurrence of a state with gapped magnetic excitations.

\section{Microscopic Realization}
In this work, we propose a way to generate a Chern-Simons term explicitly in a spin system from the fermion-spin field coupling. This topological term is obtained upon integrating out the fermions, 
as summarized in Eq. (\ref{CSfromfermion}). 
The key ingredient is to have massive Dirac fermions coupled to the spin field from the spin sector for which we propose to consider the staggered $\pi$ flux state \cite{AffMarston+Kotliar} and   
we will now describe the scheme to obtain it.

According to Eq.(\ref{tightbindingH}), the bare hopping integral is modified by the overlap of spin coherent states $t\rightarrow t\langle \Omega_{\mathbf{r}}|\Omega_{\mathbf{r}'}\rangle$.
This spin coherent state overlap generally takes complex values which gives an exponential phase factor where the argument of the exponent, to be referred to as link spin field, contains a static background flux plus the fluctuating part,
representing a dynamical spin field. In order to get Dirac spectrum, the static background flux has to take an appropriate configuration, which can be chosen to be staggered $\pi$ flux configuration \cite{AffMarston+Kotliar}.
In order to attain such staggered $\pi$ flux configuration, in turn, the spin sector should take a particular spin configuration accordingly. In the Appendix A, we first derive the expression for the link spin field
from evaluating the spin coherent state overlap explicitly. Then, from that we determine the corresponding spin configuration to obtain the staggered $\pi$ flux state.
We then propose a microscopic spin Hamiltonian that stabilizes the required spin configuration. We summarize the results as follows.

\begin{figure}
 \centering
 \includegraphics[scale=0.15]{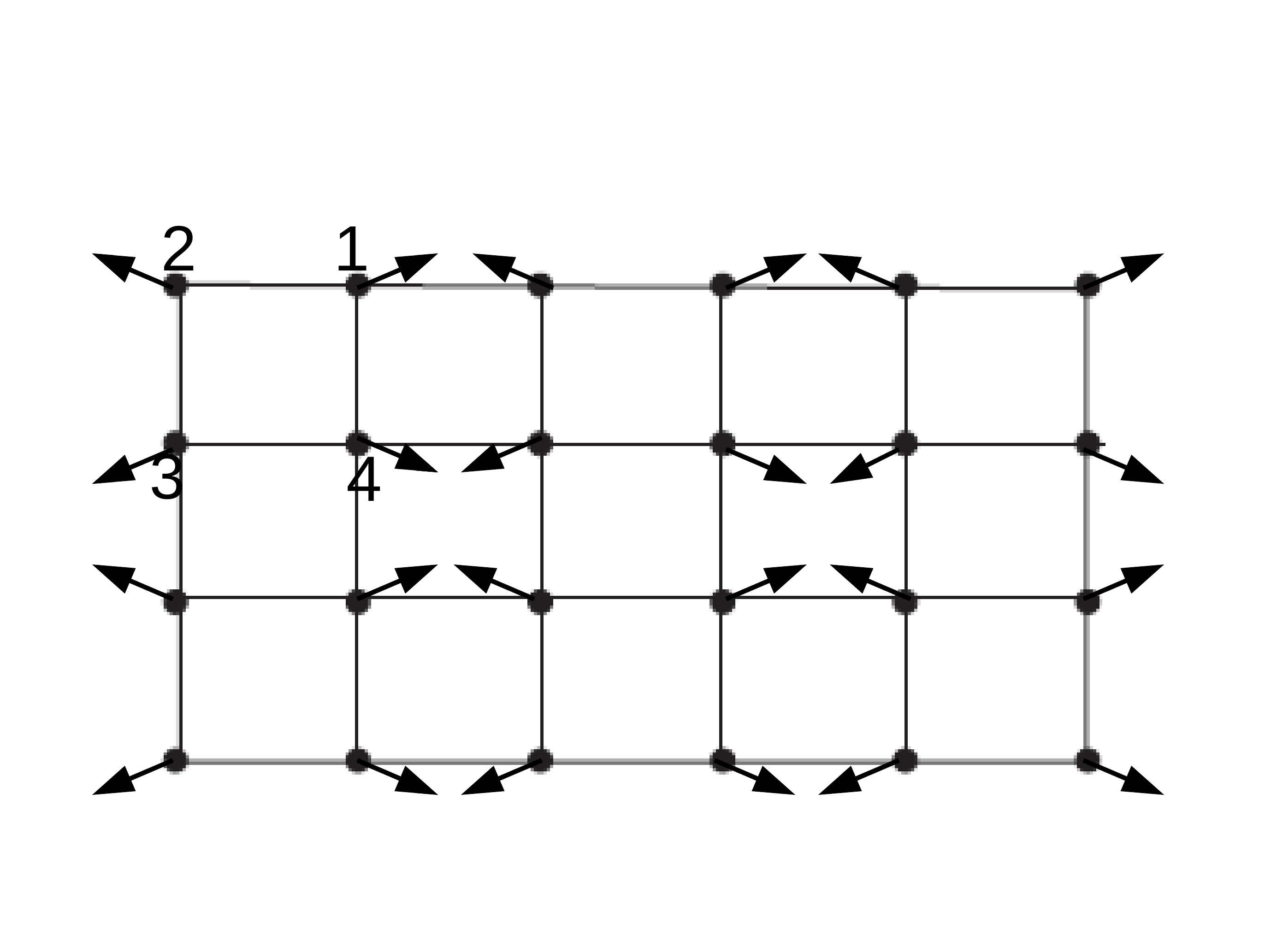}
    \caption{The planar spin configuration on square lattice with the labelling of the four sublattices.}
\label{fig:StaggeredCantedSpinConfiguration}
\end{figure}

Representing the spin as a classical vector $\mathbf{S}=S\left(\sin\theta\cos\phi,\sin\theta\sin\phi,\cos\theta\right)$ and evaluating the spin coherent state overlap in Eq.(\ref{tightbindingH}) give the following result for the link spin field \cite{SuppMaterial}, 

\begin{equation}\label{linkgaugefield}
 a_{\mathbf{r}\mathbf{r}'}=-\tan^{-1}\left[\frac{\sin(\phi_{\mathbf{r}'}-\phi_{\mathbf{r}})\sin\frac{\theta_{\mathbf{r}}}{2}\sin\frac{\theta_{\mathbf{r}'}}{2}}{\cos\frac{\theta_{\mathbf{r}}}{2}\cos\frac{\theta_{\mathbf{r}'}}{2}+\cos(\phi_{\mathbf{r}'}-\phi_{\mathbf{r}})\sin\frac{\theta_{\mathbf{r}}}{2}\sin\frac{\theta_{\mathbf{r}'}}{2}}\right]
\end{equation}
The flux $\Phi_{\Box}$ through a square plaquette is then given by $\Phi_{\Box}=\sum_{\mathbf{r}\mathbf{r}'\in\Box}a_{\mathbf{r}\mathbf{r}'}$ where the link $\mathbf{r}\mathbf{r}'$ is taken to be such that the four links traverse the square in counter-clockwise direction.
The 'mean-field part' of the link spin field needed to give $\pi$-flux state is obtained by assigning $\theta_{\mathbf{r}},\phi_{\mathbf{r}}$ their classical lowest-energy configuration values. 
It can be checked from Eq. (\ref{linkgaugefield}), that a $\pm\pi$ flux through  square plaquette can be realized by planar  spin configuration where $\theta=\pi/2$ at all sites
while the azimuthal angle takes values such that the four spins on a square rotate by a total of $\pm 2\pi$ as illustrated in Fig.(\ref{fig:StaggeredCantedSpinConfiguration}). 
The need for such planar spin configuration fully agrees with the intuition that the total solid angle $\Phi$ swept by the four spins 
around the square should equal $2\pi$ for them to produce net flux $\Phi_S=S \Phi$ equals $\pi$ with $S=1/2$. 

Similar planar spin configurations can be realized using $J_1-J_2$ model with impurity \cite{FMila}. Here we propose an alternative method. 
We find that such kind of spin configurations can be stabilized as the classical lowest-energy configuration of the following Hamiltonian, involving Dzyaloshinskii-Moriya (DM) interactions \cite{DzyaloshinskiiMoriya}.

\begin{equation}\label{latticespinmodel}
 H=J\sum_{ij} \mathbf{S}_i\cdot\mathbf{S}_j+D\sum_i(S^z_i)^2 +\sum_{ij}\mathbf{D}^{ij}_{DM}\cdot (\mathbf{S}_i\times \mathbf{S}_j)
\end{equation}
The easy-plane anisotropy ($D>0$) favors in-plane N\'{e}el state. 
To retain global $U(1)$ symmetry, only the component of $\mathbf{D}^{ij}_{DM}$ along $z$ direction can be allowed to be nonzero.
DM interaction usually occurs in lattice systems with low degree of symmetries. 
Here, we consider staggered DM interaction where ${D^z_{DM}}_{i,i\pm \hat{x}}=-{D^z_{DM}}_{i,i\pm \hat{y}}$ while ${D^x_{DM}}_{ij}={D^y_{DM}}_{ij}=0$ \cite{SuppMaterial}.
According to Moriya's symmetry analysis \cite{DzyaloshinskiiMoriya}, this DM pattern is in principle allowed as long as the system has no bond-centered inversion symmetry but 
has mirror planes perpendicular to the bonds and passing through the bond centers. 

To find the lowest energy spin configuration, we re-write the Hamiltonian Eq.(\ref{latticespinmodel}) in terms of angular variables $\theta,\phi$ as before and minimize $H$ with respect to these variables \cite{SuppMaterial}.
The lowest energy configuration of $H$ in Eq.(\ref{latticespinmodel}) is found to be a planar spin configuration with azimuthal angles $\Delta\phi_{ij}=\phi_j-\phi_i$ between nearest-neighbor spins given by 
\begin{equation}\label{dphi}
\Delta\phi_{ij}=\tan^{-1}\left(\frac{D^{ij}_{DM}}{J}\right)
\end{equation}
as shown in Fig.(\ref{fig:StaggeredCantedSpinConfiguration}). Our calculation confirms that the planar spin configuration 
is indeed the lowest-energy configuration of $H$ at least within a finite regime of phase diagram defined in parameter space ($J-D-D_{DM}$ space) and 
therefore requires no fine tuning.

We then derive from Eq.(\ref{tightbindingH}) the low-energy effective theory of fermions around the Dirac points coupled to the spin quantum fluctuations around the lowest energy spin configuration.
This is done by expanding the spin vector around its lowest energy orientation and expanding the fermion field operator using gradient expansion in real space and expansion around Dirac point in momentum space.
The details are given in Appendix B. The resulting scalar theory takes the form given in Eq.(\ref{spinsectoraction1}) for the spin sector with $\phi\equiv\phi_G=(\sum^{N_{\mathrm{unitcell}}}_{i=1}\phi_i)/N_{\mathrm{unitcell}}$ 
(where $\phi_i$ is the $\phi$ of the $i^{th}$ sublattice) representing the massless Goldstone mode. 
Here, as the other fields are gapped, gapped magnetic excitations imply spin gap.
The Berry phase term gives a contribution $\exp(i2\pi S)$ to the partition function in imaginary time. 
For integer $S$ this equals unity corresponding to constructive interference and one can thus expect the presence of spin gapped state. 
The fermion-spin field theory is given by Eq.(\ref{holeaction}) with $\psi(\mathbf{r})=(c^1_D(\mathbf{r}),c^2_D(\mathbf{r}),c^3_D(\mathbf{r}),c^4_D(\mathbf{r}))^T$.
Here $c^i_D,i=1,\cdots,4$ represents the Dirac fermion operator for the $i^{th}$ sublattice, while $\gamma_{\mu},\mu=0,1,2$ are $4\times 4$ matrices 
satisfying the Clifford algebra $\{\gamma_{\mu},\gamma_{\nu}\}=2g_{\mu\nu}$ and $\overline{\psi}=\psi^{\dag}\gamma_0$ \cite{SuppMaterial}. 
We take the unit hopping integral $t=1$ and unit lattice spacing $a=1$ as units
of energy and length respectively. One important result is that the gauge charge is found to be 

\begin{equation}\label{gaugecharge}
e=-\frac{1}{2\sqrt{2}}\left[1-\frac{J}{\sqrt{J^2+D^2_{DM}}}\right]^{\frac{1}{2}}
\end{equation}
where $\alpha=e^2/(4\pi)$ gives the dimensionless 'fine structure constant' of the $U(1)$ gauge theory.
This gauge charge is obtained directly as the coupling constant of the derived fermion-spin field theory.
It is to be noted that in realistic situations, $|D_{DM}|\ll J$ and as a result, $\alpha$ becomes a small parameter, as is normally the case in quantum field theory.
It will be shown in the next section that this fact will play a key role in the mechanism of protection against spin gap that we propose in this work.
Strictly speaking, the derivation of our fermion-spin field theory is justified only in the large spin $S$ limit.
While we have performed explicit calculation to obtain the quantitative result for the gauge charge $e$ as above using $S=1/2$ \cite{SuppMaterial} as example,
our main result regarding novel mechanism for protection against spin gap that follows soon will still hold qualitatively and apply to general spin $S$.

\begin{figure}
 \centering
 \includegraphics[scale=0.150]{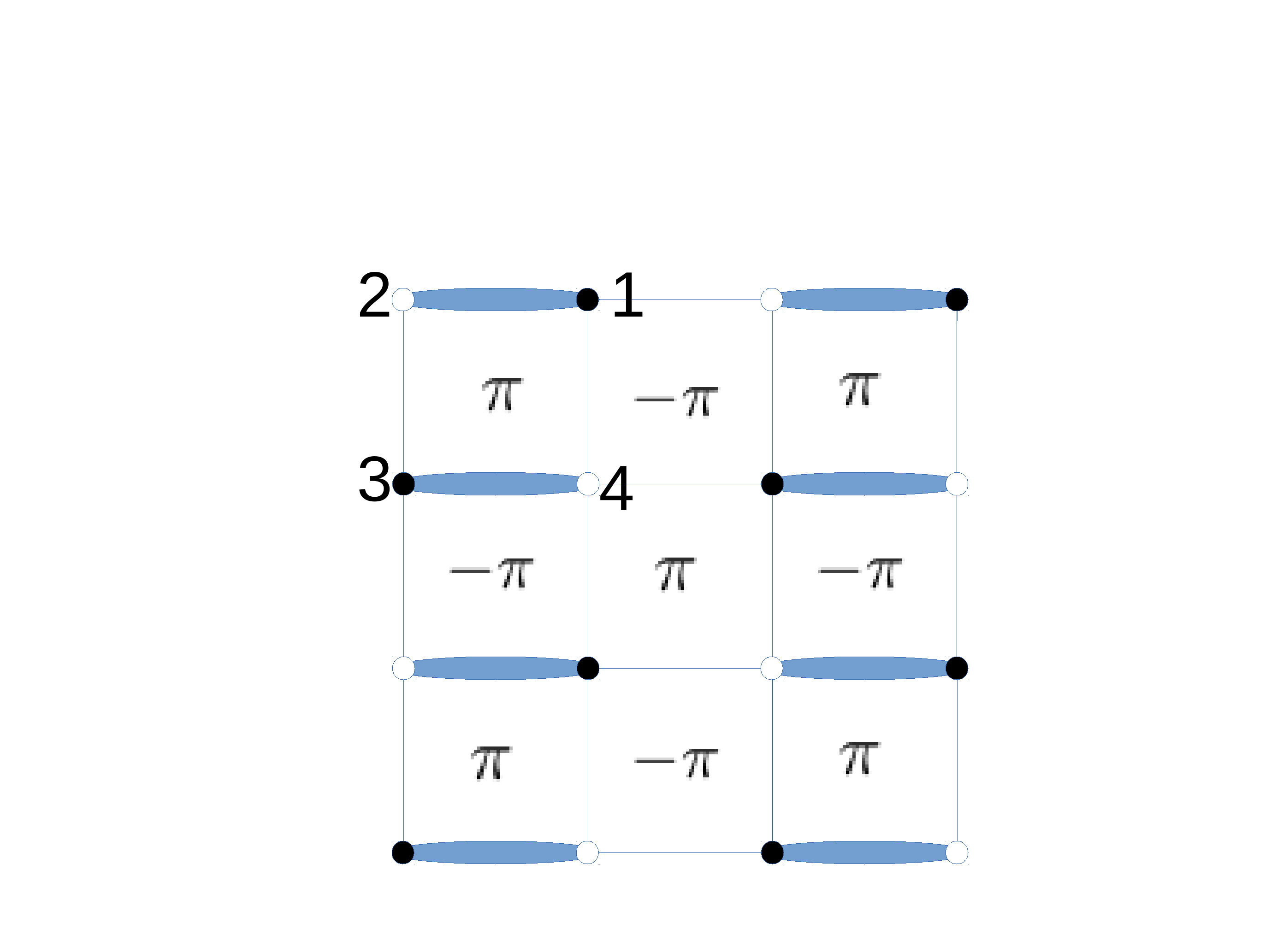}
 \caption{The columnar dimerized plus staggered chemical potential added to the staggered $\pi$ flux state.}
\label{fig:waysofgettingmass}
\end{figure}

Now we show how to get the Dirac mass for the massive Dirac fermion spectrum. We consider a perturbation on the ideal system in terms of dimerization of the strength of the hopping integral and staggerization of the sublattice chemical potential. 
Since we aim for a Chern-Simons effective theory, which intrinsically breaks discrete symmetries, the full system consisting of the spin and fermion sectors including the perturbation 
should break the symmetry under parity and time reversal times any lattice translation. Eq. (\ref{CSfromfermion}) also suggests that Dirac fermion masses from different 
fermion flavors must necessarily have the same sign in order to ensure nonzero net Chern-Simons term, as masses of opposite signs will give no net parity-breaking effect \cite{Semenoff}. 
We find that this can be achieved by considering a combination of a columnar dimerization and a staggered sublattice chemical potential
with profiles shown in Fig.(\ref{fig:waysofgettingmass}). It gives a Dirac mass Hamiltonian

\begin{equation}\label{DiracmassHamiltonian}
 H_m=\int d^2 x [\overline{\psi}'_{1}m_{\psi}\psi'_{1}+\overline{\psi}'_{2}m_{\psi}\psi'_{2}]
\end{equation}
where $\overline{\psi}'_{1(2)},\psi'_{1(2)}$ represent the first (second) pair of sublattice 1-sublattice-2 spinor $\psi'_{1(2)}=(c'_{1(3)},c'_{2(4)})^T$ in diagonalized basis, 
with $\gamma^0=\tau^z$, the Pauli matrix in this sublattice space. The resulting mass is found to be $m_{\psi}=\sqrt{m^2_{\mathrm{dimer}}+m^2_s}$ where $m_{\mathrm{dimer}}=\eta-1$ with $\eta>1$ 
is the strength of dimerized bond with respect to normal bond and $m_s$ is 
the staggered sublattice chemical potential. The above perturbation is an example but in principle, any perturbations breaking the same discrete symmetries 
are expected to give rise to the same phenomena.

\section{Discussion}

A Chern-Simons term is known to give rise to a mass to the gauge field in the Maxwell-Chern-Simons theory. What we have here is an $XY$-Chern-Simons theory plus the ever important Berry phase term. 
Plateaus in $XY$ model without a Chern-Simons term occur when the spin sector is gapped and vortex-loops proliferate corresponding to condensation of bosons. 
In the presence of the Chern-Simons term, vortex loops are anyon worldlines as we noted from Eq. (\ref{partitionfunction}), with statistical phase $\Theta=2\kappa \pi=\pi e^2/2$.
If the vortex loops could ever condense, we would realize the analog of a fractional quantum Hall state (FQHE) with its chiral edge states.
For our microscopic model with gauge charge $e$ as given in Eq. (\ref{gaugecharge}), we obtain an upper bound $\Theta=\pi/16$ from the limit $|D_{DM}|/J\rightarrow \infty$ which means that
in order to obtain condensation of anyons, we need at minimum $\mathcal{N}_{\mathrm{anyon}}=32$ vortex loops to form a boson first before they can ever condense and give rise to spin gapped state, 
taking the fact that for bosons, the statistical angle is $\Theta=2\pi$. 
Realistic situations where $|D_{DM}|/J\ll 1$ require much larger number of anyons for them to condense.

Based on the analogy with the FQHE, condensation of anyons and thus Chern-Simons-induced spin-gapped state in doped antiferromagnets can occur only when the microscopic parameters give rise to 
a Chern-Simons term with coupling $\kappa$ such that $\nu\equiv 2\kappa=\Theta/\pi=e^2/2=P/Q$ satisfies a continued fraction expansion condition, where $\nu$ is the filling factor 
of typical values $1/\nu\leq 9$ for Laughlin's states \cite{Fisher-Lee}. With a minimum of 32 vortex loops needed to form a boson (equivalent to $1/\nu\gg 16$ for the realistic case $|D_{DM}|/J\ll 1$) and the strict condition on the 
Chern-Simons coupling, it is thus in general very unlikely to form such vortex loops condensate and the associated spin gapped state. 
We can therefore conclude that the spin system Eq.(\ref{latticespinmodel}) is protected from being in a spin gapped state,
due to the Chern-Simons term induced by the fermion-spin field coupling in doped antiferromagnets with a massive Dirac fermion spectrum. In contrast, the conventional $XY$
model with the Berry phase term would mandate the occurrence of a spin gapped state. On the other hand, if the net Chern-Simons term vanishes, 
then the protection effect is inactive and one can get back a conventional spin gapped state that occurs for integer spin $S$.
It is to be noted that this result on the novel mechanism for protection against spin gap is valid for general spin $S$ because the applicability
of the mechanism is determined more by the value of coupling constants of the spin model ($J$ and $D_{DM}$) rather than the spin $S$ itself.

It would be interesting to find other microscopic spin models which can generate Chern-Simons term via similar fermion-spin field coupling as we proposed here and yet are able to
induce anyonic vortex loop condensation, giving rise to a spin gapped state, and the analog of the FQHE with chiral edge states in spin systems.
Our choice for square lattice is because a discrete Chern-Simons gauge theory with precisely the same physics as continuum one can be constructed consistently 
on this lattice \cite{Eliezer-Semenoff} as well as on kagome lattice \cite{FradkinCSkagome}, due to the presence of one-to-one face-vertex correspondence on these lattices \cite{FradkinCSlattice}. 
We would like to propose a study on compounds $\mathrm{La_2CuO_4}$ \cite{Cuprates} and
$\mathrm{LaMnO_3}$ \cite{Manganite} which are effectively 2-d square lattice antiferromagnets as the candidate materials to test our theory.
Beyond these particular materials, we claim that this novel scenario of spin gap protection is general enough to be expected anytime a Chern-like charge insulator is at play as the high energy sector. 
The Kondo lattice model in the triangular lattice may be another possible laboratory \cite{MaBa}.

\section{Acknowledgements}IM is supported by the grant No. ANR-10-LABX-0037 of the Programme des Investissements d'Avenir of France.
We thank M. Oshikawa, A. Tanaka, D. Poilblanc, and I. Affleck for the insightful discussions and C. Mudry and C. Chamon for the critical reading of the manuscript.

\appendix

\section{The Classical Lowest-Energy Spin Configuration}

\begin{figure}
 \centering
   \includegraphics[scale=0.20]{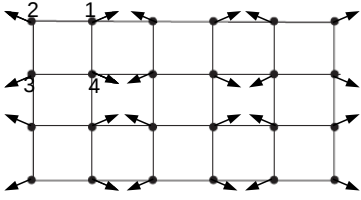}
    \caption{The (commensurate) planar spin configuration on a square plaquette of the square lattice giving rise to staggered $\pi$ flux state of Affleck-Marston ansatz.}
\label{fig:StaggeredCantedSpinConfiguration1}
\end{figure}

A key requirement in our proposal to generate Chern-Simons term in a spin system is to have Dirac spectrum for the fermions that are to be doped into the system. 
In order to get a Dirac spectrum on the square lattice, we need a $\pi$-flux state of the type as first proposed by Affleck-Marston and G. Kotliar \cite{AffMarston+Kotliar}.
The flux per square plaquette is a staggered $\pi$ and $-\pi$ configuration, where the flux alternates from $\pi$ to $-\pi$ between adjacent squares. 
Here we will show that such staggered $\pi$ flux state can be realized by a planar spin configuration illustrated in Fig.(\ref{fig:StaggeredCantedSpinConfiguration1}). 
The derivation to arrive at such spin configuration is as follows. 

The spin sector couples to the fermion by providing the spin field in the form of Goldstone modes in addition to providing the background staggered $\pi$-flux state.
This is well described by the following tight-binding Hamiltonian.

\begin{equation}\label{tightbindingH1}
 H=-t\sum_{\mathbf{r},\mathbf{r}'}\langle \Omega_{\mathbf{r}}|\Omega_{\mathbf{r}'}\rangle c^{\dag}_{\mathbf{r}'}c_{\mathbf{r}}+h.c.
\end{equation}

In spherical coordinate, the classical spin vector is written as $\mathbf{S}=S\left(\sin\theta\cos\phi,\sin\theta\sin\phi,\cos\theta\right)$.
The spin coherent state for general spin $S$ is given by
\begin{widetext}
\begin{equation}\label{spincoherentstate}
 |\Omega_{\mathbf{r}}(\theta(\mathbf{r}),\phi(\mathbf{r})\rangle=e^{iSb_{\mathbf{r}}}\sqrt{(2S)!}\sum^S_{m_z=-S}\frac{(\cos\frac{\theta_{\mathbf{r}}}{2}e^{i\frac{\phi_{\mathbf{r}}}{2}})^{S+m_z}(\sin\frac{\theta_{\mathbf{r}}}{2}e^{-i\frac{\phi_{\mathbf{r}}}{2}})^{S-m_z}}{\sqrt{(S+m_z)!(S-m_z)!}}|S,m_z\rangle
\end{equation}
\end{widetext}
which for $S=1/2$ allows us to write

\begin{equation}\label{coherentstate}
 |\Omega_{\mathbf{r}}\rangle=e^{i\frac{1}{2} b_{\mathbf{r}}}\left(e^{i\frac{\phi_{\mathbf{r}}}{2}}\cos\frac{\theta_{\mathbf{r}}}{2},e^{-i\frac{\phi_{\mathbf{r}}}{2}}\sin\frac{\theta_{\mathbf{r}}}{2}\right)^T
\end{equation}
where $b_{\mathbf{r}}$ is a pure gauge function which we can set to constant function $b_{\mathbf{r}}=b_0$ in the simplest case.
The coherent state overlap for electron where $S=1/2$ is given by
\begin{widetext}
 \begin{equation}\label{overlap}
\langle \Omega_{\mathbf{r}}|\Omega_{\mathbf{r}'}\rangle=e^{\frac{1}{2}i(b_{\mathbf{r}'}+\phi_{\mathbf{r}'}-b_{\mathbf{r}}-\phi_{\mathbf{r}})} \left(\cos\frac{\theta_{\mathbf{r}}}{2}\cos\frac{\theta_{\mathbf{r}'}}{2}+e^{i(\phi_{\mathbf{r}}-\phi_{\mathbf{r}'})}\sin\frac{\theta_{\mathbf{r}}}{2}\sin\frac{\theta_{\mathbf{r}'}}{2}\right)
\end{equation}
\end{widetext}
This spin coherent state overlap gives rise to a spin field as follows. Taking the gauge function to be $b_{\mathbf{r}}=-\phi_{\mathbf{r}}$,

\begin{equation}\label{linkgaugefield1}
 a_{\mathbf{r}\mathbf{r}'}=-\tan^{-1}\left[\frac{\sin(\phi_{\mathbf{r}'}-\phi_{\mathbf{r}})\sin\frac{\theta_{\mathbf{r}}}{2}\sin\frac{\theta_{\mathbf{r}'}}{2}}{\cos\frac{\theta_{\mathbf{r}}}{2}\cos\frac{\theta_{\mathbf{r}'}}{2}+\cos(\phi_{\mathbf{r}'}-\phi_{\mathbf{r}})\sin\frac{\theta_{\mathbf{r}}}{2}\sin\frac{\theta_{\mathbf{r}'}}{2}}\right]
\end{equation}
We will consider the Dirac spectrum obtained from the staggered $\pi$ flux state configuration of Affleck-Marston. 
It can be checked from Eq. (\ref{linkgaugefield1}) that for $\theta_{\mathbf{r}}=\pi/2$ at all sites $\mathbf{r},\mathbf{r}'$, we will always get $\sum_{\mathbf{r}\mathbf{r}'\in\Box}a_{\mathbf{r}\mathbf{r}'}=\pm\pi$, 
as the link spin field is then given by $a_{\mathbf{r}\mathbf{r}'}=-\Delta\phi_{\mathbf{r}\mathbf{r}'}/2$ where $\Delta\phi_{\mathbf{r}\mathbf{r}'}=\phi_{\mathbf{r}'}-\phi_{\mathbf{r}}$. 
This gives rise to the needed $\pi$ flux per square plaquette. 
This remarkable result will give rise to the planar  spin configuration shown in Fig.(\ref{fig:StaggeredCantedSpinConfiguration1}) for an appropriate choice of Hamiltonian. 

We will prove here that this planar spin configuration can be obtained as the lowest energy configuration of the following Hamiltonian,

\begin{equation}\label{latticespinmodel1}
 H=J\sum_{ij} \mathbf{S}_i\cdot\mathbf{S}_j +D\sum_i (S^z_i)^2 +\sum_{ij}\mathbf{D}^{ij}_{DM}\cdot (\mathbf{S}_i\times \mathbf{S}_j)
\end{equation}
consisting of a Heisenberg antiferromagnetic coupling ($J>0$), easy-plane anisotropy ($D\geq 0$), and a Dzyaloshinskii-Moriya terms. 
To retain global $U(1)$ symmetry, only one ($x,y$ or $z$) component of $D^{ij}_{DM}$ normal to easy plane can be allowed to be nonzero; for concreteness we choose $z$ component.
It will be shown here that in order to get the planar  spin configuration Fig. (\ref{fig:StaggeredCantedSpinConfiguration1}),
we must take ${D^{ij}_{DM}}^z>0$ on $x$ links and ${D^{ij}_{DM}}^z<0$ on $y$ links (of equal magnitude but opposite in sign) or vice versa.

In spherical coordinate, the full Hamiltonian is given by
\begin{widetext}
\begin{equation}
H=JS^2\sum_{ij}[\sin\theta_i\sin\theta_j\cos(\phi_j-\phi_i)+\cos\theta_i\cos\theta_j] +DS^2\sum_i\cos^2\theta_i +\sum_{ij}D^{ij}_{DMz}S^2\sin\theta_i\sin\theta_j\sin(\phi_j-\phi_i)
\end{equation}
\end{widetext}

To find the lowest-energy configuration, we take $\partial H/\partial\phi_i=0,\partial H/\partial\theta_j=0$ and verify whether $\partial^2H/\partial\phi_i\partial\phi_j>0,\partial^2H/\partial\theta_i\partial\theta_j>0,\partial^2H/\partial\phi_i\partial\theta_j=0$.
We find the lowest energy configuration solution where the nearest-neighbor spins have polar angles as $\theta_i=\pi/2$ at all sites $i$ and azimuthal angle $\Delta\phi_{ij}=\phi_j-\phi_i$ between nearest-neighbor spins given by 
\begin{equation}\label{dphis}
\Delta\phi_{ij}=\tan^{-1}\left(\frac{D^{ij}_{DM}}{J}\right)
\end{equation}
whereas next-nearest neighbor spins across the diagonal of square have $\Delta\phi_{ij}=\phi_j-\phi_i=\pm\pi$ as shown in 
Fig.(\ref{fig:StaggeredCantedSpinConfiguration1}). We verify that the conditions on the second derivatives are indeed satisfied. 
The above calculation guarantees that such planar  spin state indeed exists within a finite regime of the phase diagram defined 
in the parameter space ($J\comment{-D}-D-D_{DM}$ space) and therefore requires no fine tuning.

Now we can describe the precise resulting lowest energy spin configuration for a particular pattern of the Dzyaloshinskii-Moriya interaction coupling $\mathbf{D}^{ij}_{DM}$ of relevance to our purpose.
For case 1), we take $\mathbf{D}^{ij}_{DM}=\mathbf{D}_{DM}$ to be uniform equal to a constant vector at all links, and without loss of generality we take it to point to the $z$ direction for simplicity.
It can be checked from Eq. (\ref{dphis}) that unless $|\mathbf{D}_{DM}|/J=1$ giving $\Delta\phi_{ij}=\pi/4,5\pi/4$, one cannot get $\pi$ flux on a square plaquette for realistic case where $|\mathbf{D}_{DM}|/J\ll 1$.
The four spins just cannot be arranged to rotate in total by $2\pi$ and at the same satisfy Eq. (\ref{dphis}) from link to link on a square. What this means is that we have to consider a plaquatte involving
further distanced spins to get $\pi$ flux. Thus in general we will get an incommensurate state for case 1). For case 2), we consider ${\mathbf{D}_{DM}}_{i,i\pm \hat{x}}=(0,\pm d_1,0)$ on each 'column' and alternating ${\mathbf{D}_{DM}}_{i,i\pm \hat{y}}=(0,\pm d_2,0)$ on each 'row'
of square lattice, where $|d_1|=|d_2|$ and so $d_1=\pm d_2$. It can be easily checked using Eq. (\ref{dphis}) that one can arrange the four spins to rotate by $2\pi$ on a square but when we tile up the pattern on a lattice,
it will take a relatively large cluster of squares with their spins to form unit cell from which the full lattice spin configuration can be formed. We however get a commensurate staggered $\pi$ flux state.
This DM pattern occurs in real materials, e.g. $\mathrm{La_2CuO_4}$ \cite{Cuprates} but such large unit cell complicates the the derivation of Dirac fermion field theory as the spinor and matrix sizes grow with unit cell size.
We then consider the model case 3) where we have ${D^z_{DM}}_{i,i\pm \hat{x}}=-{D^z_{DM}}_{i,i\pm \hat{y}}$ while ${D^x_{DM}}_{ij}={D^y_{DM}}_{ij}=0$. It can be checked using Eq. (\ref{dphis}) that the four spins on a square
will take configuration as shown in Fig. (\ref{fig:StaggeredCantedSpinConfiguration1}) and this can be nicely repeated over the whole square lattice and gives what we call (commensurate) planar  spin configuration
with the resulting commensurate staggered $\pi$ flux state with just four sublattices (four sites per unit cell). The needed DM pattern is in principle allowed by symmetry according to Moriya's original consideration. 
We take this latter case as example but our final results apply to the real materials mentioned above. 

\begin{figure}
 \centering
  \includegraphics[scale=0.150]{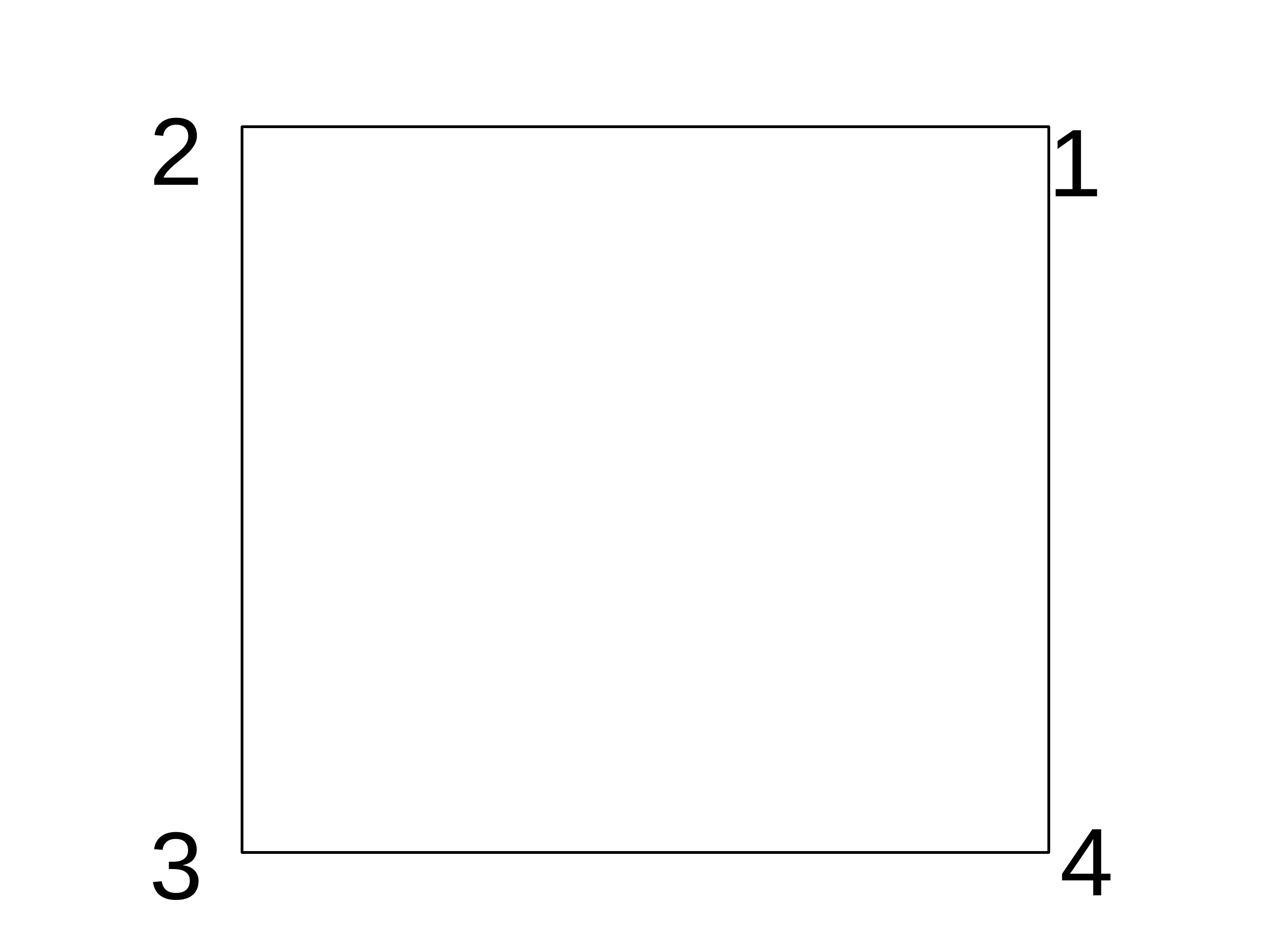}
   \caption{Labeling of the four sublattices is as shown in the square.
   The labeling of the four Dirac points in Brillouin zone also follows the same numbering.}
\label{fig:Labeling}
\end{figure}
We label the four sublattices as shown in Fig.(\ref{fig:Labeling}).
It is to be noted that in such planar spin configuration, each sublattice has its own $\theta$ and $\phi$.
We therefore have four sets of $(\phi^{(i)},\theta^{(i)})$ fields: $(\phi^{(1)},\theta^{(1)})$,$(\phi^{(2)},\theta^{(2)})$,$(\phi^{(3)},\theta^{(3)})$,$(\phi^{(4)},\theta^{(4)})$.
From these, we can define four orthogonal fields;

\[
 \phi_G=\frac{1}{4}\left(\phi_1+\phi_2+\phi_3+\phi_4\right),\\
 \phi_{t1}=\frac{1}{4}\left(\phi_1+\phi_2-\phi_3-\phi_4\right),\\
 \]
 \begin{equation}
 \phi_{t2}=\frac{1}{4}\left(\phi_1-\phi_2+\phi_3-\phi_4\right),\\
 \phi_{t3}=\frac{1}{4}\left(\phi_1-\phi_2-\phi_3+\phi_4\right) 
\end{equation}
\[
 \theta_G=\frac{1}{4}\left(\theta_1+\theta_2+\theta_3+\theta_4\right),\\
 \theta_{t1}=\frac{1}{4}\left(\theta_1+\theta_2-\theta_3-\theta_4\right),\\
\]
 \begin{equation}
 \theta_{t2}=\frac{1}{4}\left(\theta_1-\theta_2+\theta_3-\theta_4\right),\\
 \theta_{t3}=\frac{1}{4}\left(\theta_1-\theta_2-\theta_3+\theta_4\right) 
\end{equation}
and the inverse mappings
\[
 \phi_1=\left(\phi_G+\phi_{t1}+\phi_{t2}+\phi_{t3}\right),\\
 \phi_2=\left(\phi_G+\phi_{t1}-\phi_{t2}-\phi_{t3}\right),\\
\]
\begin{equation}
 \phi_3=\left(\phi_G-\phi_{t1}+\phi_{t2}-\phi_{t3}\right),\\
 \phi_4=\left(\phi_G-\phi_{t1}-\phi_{t2}+\phi_{t3}\right) 
\end{equation}
\[
 \theta_1=\left(\theta_G+\theta_{t1}+\theta_{t2}+\theta_{t3}\right),\\
 \theta_2=\left(\theta_G+\theta_{t1}-\theta_{t2}-\theta_{t3}\right),\\
\]
\begin{equation}
 \theta_3=\left(\theta_G-\theta_{t1}+\theta_{t2}-\theta_{t3}\right),\\
 \theta_4=\left(\theta_G-\theta_{t1}-\theta_{t2}+\theta_{t3}\right) 
\end{equation}
The effective low energy theory of the spin sector from the Hamiltonian Eq. (\ref{latticespinmodel1}) will give rise to a gapless Lagrangian for $\phi_G$ and gapped ones for
$\phi_{ti},\theta_{ti},i=1,2,3$, so that these latter fields can all be integrated out in the low energy physics \cite{LCP}. We therefore have only the $\phi_G$ remaining in the low energy 
physics. This $\phi_G$ is nothing but the Goldstone modes of the rotationl symmetry-breaking classical lowest-energy configuration that we have obtained in the beginning: the
planar  spin configuration. The effective low energy description of the spin sector obtained upon integrating out all massive fields take the form

\begin{equation}
 \mathcal{L}=\frac{K_{\mu}}{2}\partial^{\mu}\phi_G\partial_{\mu}\phi_G+i\frac{S}{a^2}\partial_{\tau}\phi_G
\end{equation}
and it is to be noted that the Berry phase term is obtained from summing over the Berry phases of the four sublattices
which finally involves only the diagonal field $\phi_G$ (Goldstone mode) with overall prefactor $S$.
Since upon integration over $\tau$ Berry phase gives phase factor $\exp(i2\pi S)$, it equals unity for integer spin $S$
and thus has no effect in such case. To simplify notation, in the rest of this calculation, $\phi,\theta$ represent $\phi_G,\theta_G$.



\section{Derivation of Effective Fermion-Spin Field Action of Insulating Antiferromagnet with Massive Dirac Spectrum}

We have shown in the previous section the microscopic spin model to obtain (gapless) Dirac spectrum for the fermions doped into the system. 
We have to verify that the resulting low-energy effective theory is indeed that of Dirac fermion. Furthermore, in order to generate the Chern-Simons term 
properly, the Dirac fermions must be massive rather than massless. We will therefore also demonstrate in this section the way to generate
massive Dirac fermion spectrum.

\begin{figure}
 \centering
 \includegraphics[scale=0.20]{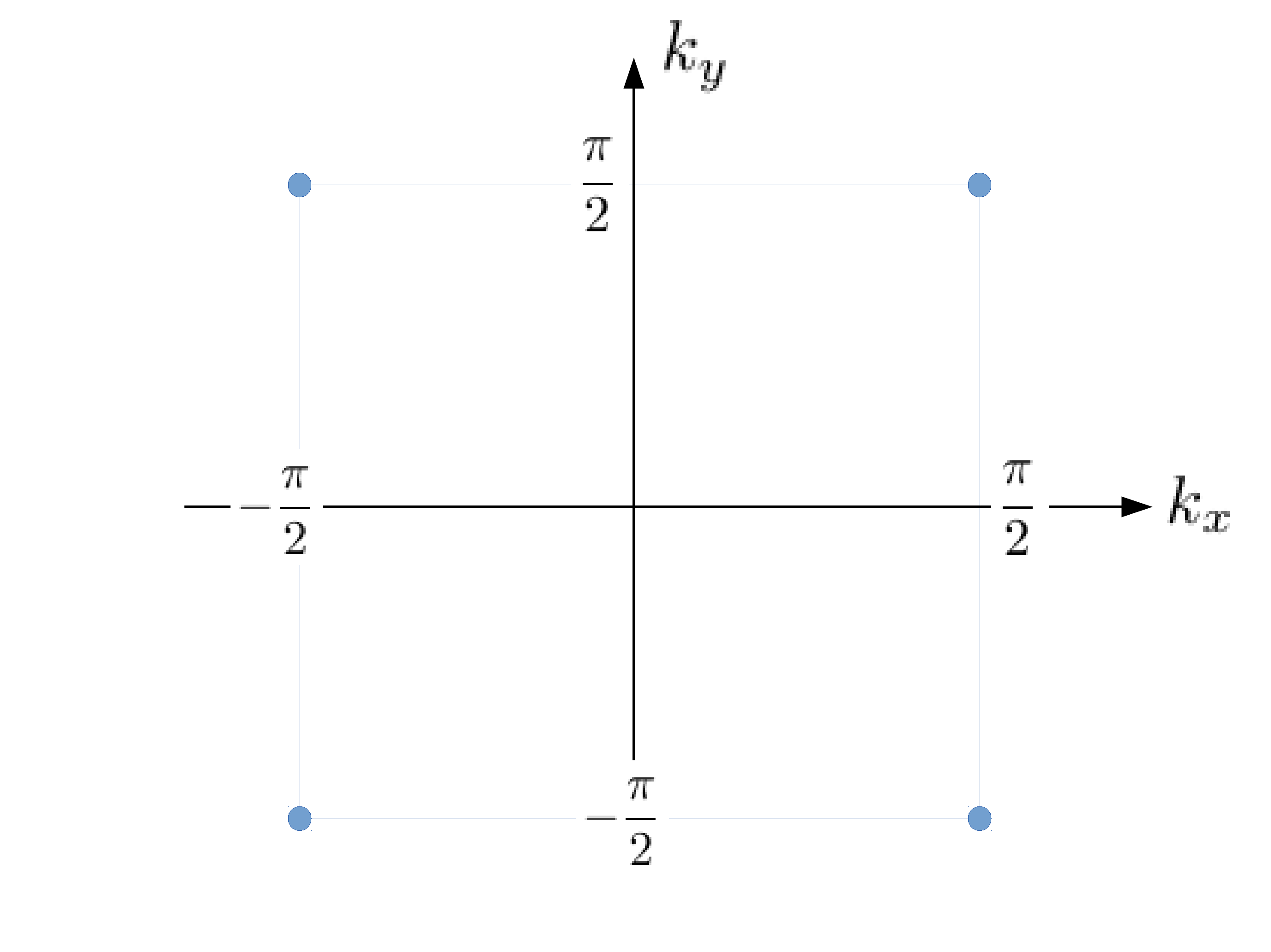}
 \caption{The $1^{st}$ Brillouin zone of the staggered $\pi$ flux state in the 4-sublattice desciption and the Dirac points (solid circles).}
\label{fig:1BZDiracPoints4Sublattice}
\end{figure}

To derive the low-energy effective theory of fermions coupled to the spin field, we start from Eq. (\ref{tightbindingH1}). 
Fermion in the staggered $\pi$-flux state of Affleck-Marston has Dirac spectrum $E_k=\pm 2\chi_1\sqrt{\cos^2k_x+\cos^2k_y}$ with 
four Dirac points at $\mathbf{k}=(\pm \pi/2,\pm \pi/2)$ as shown in Fig.(\ref{fig:1BZDiracPoints4Sublattice}) with their numbering.
In the 4-sublattice description of Affleck-Marston $\pi$-flux state, the new first Brillouin zone is defined by the square with $k_x, k_y=\pm \pi/2$.

We first expand the fermion field operator around these four Dirac points

\[
 c(\mathbf{r})=\int \frac{d^2k}{(2\pi)^2}c_{\mathbf{k}}e^{i\mathbf{k}\cdot\mathbf{r}}\approx \sum^4_{i=1}\int_{\mathbf{k}\approx \mathbf{k}_{Di}} \frac{d^2k}{(2\pi)^2}c_{\mathbf{k}}e^{i\mathbf{k}\cdot\mathbf{r}}
\]
\begin{equation}
 =\sum^4_{a=1}e^{i\mathbf{k}_{Da}\cdot\mathbf{r}}c_{Da}(\mathbf{r})
\end{equation}
where the slow-fermion field operator $c_{Da}(\mathbf{r})$ is given by

\[             
 c_{Da}(\mathbf{r})=\int_{\mathbf{k}\approx \mathbf{k}_{Da}} \frac{d^2k}{(2\pi)^2}c_{\mathbf{k}}e^{i(\mathbf{k}-\mathbf{k}_{Da})\cdot\mathbf{r}}
\]
\begin{equation}
 \approx \int_{|\mathbf{q}_a|\leq q_c} \frac{d^2q}{(2\pi)^2}c_{\mathbf{k}_{Da}+\mathbf{q}_a}e^{i\mathbf{q}_a\cdot\mathbf{r}}
\end{equation}
with $\mathbf{q}_a=\mathbf{k}-\mathbf{k}_{Da}$ and $\mathbf{k}_{Da}$ are the momentum of the $a^{th}$ Dirac points $\mathbf{k}_{Da}=(\pm\pi/2,\pm\pi/2)$ and $q_c$ is some appropriate UV cutoff.

We then perform gradient expansion and in doing so,
we expand the spin coherent state parameterized in terms of these two angles around the respective classical lowest-energy configuration value $\theta_0$ and $\phi_0$ 
at each site (sublattice). Eventually, only the Goldstone mode $\phi_G$ survives and in the following, sublattice index is omitted and in the final results $\phi,\theta$ represent
$\phi_G,\theta_G$.
\begin{widetext}
\begin{equation}\label{cohstateexpansion1}
  |\Omega_{\mathbf{r}}(\theta_{\mathbf{r}},\phi_{\mathbf{r}})\rangle=|\Omega_{\mathbf{r}}(\theta^0_{\mathbf{r}},\phi^0_{\mathbf{r}})\rangle+\delta\theta_{\mathbf{r}}\partial_{\theta}|\Omega_{\mathbf{r}}(\theta^0_{\mathbf{r}},\phi^0_{\mathbf{r}})\rangle+\delta\phi_{\mathbf{r}}\partial_{\phi}|\Omega_{\mathbf{r}}(\theta^0_{\mathbf{r}},\phi^0_{\mathbf{r}})\rangle+\mathcal{O}(\delta\theta^2,\delta\phi^2,\delta\theta\delta\phi)
\end{equation}
\begin{equation}\label{cohstateexpansion2}
  |\Omega_{\mathbf{r}'}(\theta_{\mathbf{r}'},\phi_{\mathbf{r}'})\rangle=|\Omega_{\mathbf{r}'}(\theta^0_{\mathbf{r}'},\phi^0_{\mathbf{r}'})\rangle+\delta\theta_{\mathbf{r}'}\partial_{\theta}|\Omega_{\mathbf{r}'}(\theta^0_{\mathbf{r}'},\phi^0_{\mathbf{r}'})\rangle+\delta\phi_{\mathbf{r}'}\partial_{\phi}|\Omega_{\mathbf{r}'}(\theta^0_{\mathbf{r}'},\phi^0_{\mathbf{r}'})\rangle+\mathcal{O}(\delta\theta^2,\delta\phi^2,\delta\theta\delta\phi)
\end{equation}
\end{widetext}
\begin{equation}\label{expansion}
  c^{\dag}_{Da}(\mathbf{r}')=c^{\dag}_{Da}(\mathbf{r})+\Delta\mathbf{r}\cdot\nabla c^{\dag}_{Da}+\mathcal{O}(\Delta\mathbf{r}^2)
\end{equation}
where $\Delta\mathbf{r}=\mathbf{r}'-\mathbf{r}$, we obtain from Eq. (\ref{tightbindingH1}) 
\begin{widetext}
\[
 H=-t\sum_{\mathbf{r},\mathbf{r}'}\sum^4_{a,b=1}e^{i(\mathbf{k}_{Db}\cdot\mathbf{r}-\mathbf{k}_{Da}\cdot\mathbf{r}')}[c^{\dag}_{Da}(\mathbf{r})\langle \Omega^0_{\mathbf{r}}|\Omega^0_{\mathbf{r}'}\rangle c_{Db}(\mathbf{r})-c^{\dag}_{Da}i\Delta\mathbf{r}\cdot(-i\langle \Omega^0_{\mathbf{r}}|\Omega^0_{\mathbf{r}'}\rangle\nabla-e\mathbf{A})c_{Db}(\mathbf{r})]+h.c.
\]
\begin{equation}
 =-t\sum_{\mathbf{r}}\sum^4_{a=1}e^{-i\mathbf{k}_{Da}\cdot\Delta\mathbf{r}}[c^{\dag}_{Da}(\mathbf{r})\langle \Omega^0_{\mathbf{r}}|\Omega^0_{\mathbf{r}+\Delta\mathbf{r}}\rangle c_{Da}(\mathbf{r})-c^{\dag}_{Da}i\Delta\mathbf{r}\cdot(-i\langle \Omega^0_{\mathbf{r}}|\Omega^0_{\mathbf{r}+\Delta\mathbf{r}}\rangle\nabla-e\mathbf{A})c_{Da}(\mathbf{r})]+h.c.
\end{equation}
\begin{equation}
 =-t\sum_{\mathbf{r}}\sum^4_{a=1}\sum^4_{m,n=1}e^{-i\mathbf{k}_{Da}\cdot\Delta\mathbf{r}}[{c^m}^{\dag}_{Da}(\mathbf{r})\langle \Omega^0_{\mathbf{r}}|\Omega^0_{\mathbf{r}+\Delta\mathbf{r}}\rangle c^n_{Da}(\mathbf{r})-{c^m}^{\dag}_{Da}i\Delta\mathbf{r}\cdot(-i\langle \Omega^0_{\mathbf{r}}|\Omega^0_{\mathbf{r}+\Delta\mathbf{r}}\rangle\nabla-e\mathbf{A})c^n_{Da}(\mathbf{r})]+h.c.
\end{equation}
\end{widetext}
where the superscript $m,n=1,\cdots,4$ represents sublattice index of the spin sector and $Da=1,\cdots,4$ the Dirac point index. In what remains, we take the unit hopping integral $t=1.0$ and the unit lattice spacing $a=1.0$ 
as the units of energy and length respectively. The spin field is given by 

\begin{widetext}
\begin{equation}\label{gaugefieldA}
 e\mathbf{A}=-i\left(\delta\theta_{\mathbf{r}'}\langle\Omega^0_{\mathbf{r}}|(\partial_{\theta}|\Omega^0_{\mathbf{r}'}\rangle)+\delta\theta_{\mathbf{r}}(\partial_{\theta}\langle\Omega^0_{\mathbf{r}}|)|\Omega^0_{\mathbf{r}'}\rangle+\delta\phi_{\mathbf{r}'}\langle\Omega^0_{\mathbf{r}}|(\partial_{\phi}|\Omega^0_{\mathbf{r}'}\rangle)+\delta\phi_{\mathbf{r}}(\partial_{\phi}\langle\Omega^0_{\mathbf{r}}|)|\Omega^0_{\mathbf{r}'}\rangle\right)
\end{equation}
\end{widetext}
where in the continuum limit we do gradient expansion by defining $\nabla\theta=\delta\theta_{\mathbf{r}'}-\delta\theta_{\mathbf{r}},\nabla\phi=\delta\phi_{\mathbf{r}'}-\delta\phi_{\mathbf{r}}$. 
We have included the gauge charge $e$ coupling the spin field to the fermions. The expression Eq. (\ref{gaugefieldA}) suggests that this spin field is a Berry connection-type of pseudo-gauge field.
The actual contribution to the spin field $\mathbf{A}$ in Eq. (\ref{gaugefieldA}) only comes from the $\phi$ field part\comment{\cite{thetafieldgaugefield}}, 
corresponding to Goldstone mode representing symmetric combination of the sublattice phase fields $\phi_G=(\phi_1+\phi_2+\phi_3+\phi_4)/4$, whereas the terms containing $\delta\theta_{\mathbf{r}},\delta\theta_{\mathbf{r}'}$
give rise to massive momentum operator and correspond to the massive $\theta$ field, which can therefore be integrated out in the low energy physics. 

For $S=1/2$, we get the result for the spin field

\begin{equation}\label{gaugefieldexpression}
e\mathbf{A}(\mathbf{r})=-\frac{1}{2}\sin\theta_0\sin\frac{\Delta\phi_{\mathbf{r},\mathbf{r}+\Delta\mathbf{r}}}{2}\nabla\phi(\mathbf{r})
\end{equation}
where $\theta_0=\pi/2$.
The above result can be derived in a more direct and straightforward way by evaluating the coherent state overlap
and matching the complex phase factor exponent with spin field\comment{without performing the expansions done in Eqs. (\ref{cohstateexpansion1}) and (\ref{cohstateexpansion2})}.

\begin{equation}
\langle \Omega_{\mathbf{r}}| \Omega_{\mathbf{r}'}\rangle\sim e^{ia_{\mathbf{r}\mathbf{r}'}}
\end{equation}
and as one takes the continuum limit, one has $\mathbf{a}_{\mathbf{r}\mathbf{r}'}\rightarrow \mathbf{A}(\mathbf{r})$.
We obtain
\begin{widetext}
\begin{equation}\label{overlapresult}
 \langle \Omega_{\mathbf{r}}| \Omega_{\mathbf{r}'}\rangle=|\langle \Omega^0_{\mathbf{r}}| \Omega^0_{\mathbf{r}'}\rangle|e^{i\left(\frac{\phi_{\mathbf{r}'}+b_{\mathbf{r}'}-\phi_{\mathbf{r}}-b_{\mathbf{r}}}{2}\right)-i\tan^{-1}[\frac{\sin\Delta\phi^0_{\mathbf{r}\mathbf{r}'}\sin\frac{\theta^0_{\mathbf{r}}}{2}\sin\frac{\theta^0_{\mathbf{r}'}}{2}}{\cos\frac{\theta^0_{\mathbf{r}}}{2}\cos\frac{\theta^0_{\mathbf{r}'}}{2}+\cos\Delta\phi^0_{\mathbf{r}\mathbf{r}'}\sin\frac{\theta_{\mathbf{r}}}{2}\sin\frac{\theta^0_{\mathbf{r}'}}{2}}]-i\frac{1}{2}\sin\frac{\Delta\phi^0_{\mathbf{r}\mathbf{r}'}}{2}\sin\theta_0\nabla\phi+i\frac{a}{\sin^2\theta_0}\tan\frac{\Delta\phi^0_{\mathbf{r}\mathbf{r}'}}{2}(\Pi_{\mathbf{r}}+\Pi_{\mathbf{r}'})}
\end{equation}
\end{widetext}
where $\Delta\phi^0_{\mathbf{r}\mathbf{r}'}=\phi_{\mathbf{r}'}-\phi_{\mathbf{r}}$ and 

\begin{equation}
 \langle \Omega^0_{\mathbf{r}}|\Omega^0_{\mathbf{r}'}\rangle=\sin\theta_0\cos\frac{\Delta\phi_{\mathbf{r}\mathbf{r}'}}{2}
\end{equation}
where $\theta_0=\pi/2$ for our planar spin configuration. We can immediately identify
\begin{equation}\label{gaugefieldAalternative}
e\mathbf{A}(\mathbf{r})=-\frac{1}{2}\sin\theta_0\sin\frac{\Delta\phi^0_{\mathbf{r},\mathbf{r}+\Delta\mathbf{r}}}{2}\nabla\phi(\mathbf{r})
\end{equation}
in complete agreement with Eq. (\ref{gaugefieldexpression}), while the last term in Eq. (\ref{overlapresult}) is dependent on the momentum operator 

\begin{equation}
 \Pi_{\mathbf{r}}=-\frac{1}{2}[\delta\theta_{\mathbf{r}}\sin\theta_{\mathbf{r}}+\frac{1}{2}\delta\theta^2_{\mathbf{r}}\cos\theta_{\mathbf{r}}]
\end{equation}
which is massive and therefore in the low energy physics can be integrated out to give simply a renormalization correction
to the coefficients (couplings) in the effective fermion-spin field action without changing the physics.
It is to be noted that the expression for spin field Eq.(\ref{gaugefieldAalternative}) is proportional to the gradient of the phase field, as we expected.

Following our definition $A_{\mu}=\partial_{\mu}\phi$ and using Eqs. (\ref{dphis}) and (\ref{gaugefieldAalternative}), the above result gives as the effective gauge coupling 

\begin{equation}\label{gaugecoupling1}
e=-\frac{1}{2\sqrt{2}}\left[\left(1-\frac{J}{\sqrt{J^2+D^2_{DM}}}\right)\right]^{\frac{1}{2}}
\end{equation}
We note that if we set the Dzyaloshinskii-Moriya interaction to zero $D_{DM}\rightarrow 0$, the whole picture breaks down and the gauge charge equals zero. 
So, DM interaction is crucially needed here.

As can be seen from Fig.(\ref{fig:1BZDiracPoints4Sublattice}), the Dirac points $\mathbf{k}_{Da}=(\pm\pi/2,\pm\pi/2)$ are located precisely at the corners of the $1^{st}$ Brillouin zone.
The implication of this is that the four Dirac points are now equivalent to each other and it suffices to consider only one Dirac point, e.g. the $\mathbf{k}_{D1}=(\pi/2,\pi/2)$ Dirac point.
In the low energy limit, we define a Dirac 4-spinor associated with this representative Dirac point
$\psi(\mathbf{r})=\left(c^1_D(\mathbf{r}),c^2_D(\mathbf{r}),c^3_D(\mathbf{r}),c^4_D(\mathbf{r})\right)^T$. Here, $c^i_D(\mathbf{r}),i=1,\cdots,4$ represents the Dirac fermion annihilation operator 
for the $i^{th}$ sublattice of the spin sector in the planar configuration Fig.(\ref{fig:StaggeredCantedSpinConfiguration1}). The low-energy Hamiltonian is

\begin{equation}
 H=\int d^2x \overline{\psi}\gamma^{\mu}(-i\partial_{\mu}-eA_{\mu})\psi
\end{equation}
where $\mu=t,x,y\equiv 0,1,2$ and $\overline{\psi}=\psi^{\dag}\gamma^0$.

The next task is to find the $\gamma$ matrices. We obtain the following result.

\begin{equation}\label{Gx}
\gamma^x=i\gamma^t\left( \begin{array}{cccc}
0&\gamma^x_{12}&0&0\\
\gamma^x_{21}&0&0&0\\
0&0&0&\gamma^x_{34}\\
0&0&\gamma^x_{43}&0
\end{array} \right)
\end{equation}
\begin{equation}\label{Gx}
\gamma^y=i\gamma^t\left( \begin{array}{cccc}
0&0&0&\gamma^y_{14}\\
0&0&\gamma^y_{23}&0\\
0&\gamma^y_{32}&0&0\\
\gamma^y_{41}&0&0&0
\end{array} \right)
\end{equation}
where the matrix elements are given as follows.
\[
 \gamma^x_{12}=ite^{-ik_{D1x}}\langle\Omega^0_{\mathbf{r}2}|\Omega^0_{\mathbf{r}'1}\rangle=(\gamma^x_{21})^{\dag},\\
 \]
 \[
 \gamma^y_{14}=ite^{-ik_{D1y}}\langle\Omega^0_{\mathbf{r}4}|\Omega^0_{\mathbf{r}'1}\rangle=(\gamma^y_{41})^{\dag},\\
 \]
 \[
 \gamma^y_{23}=ite^{-ik_{D1y}}\langle\Omega^0_{\mathbf{r}3}|\Omega^0_{\mathbf{r}'2}\rangle=(\gamma^y_{32})^{\dag},\\
 \]
 \begin{equation}\label{gammamatrixelements}
 \gamma^x_{34}=ite^{ik_{D1x}}\langle\Omega^0_{\mathbf{r}4}|\Omega^0_{\mathbf{r}'3}\rangle=(\gamma^x_{43})^{\dag}
 \end{equation}
These matrix elements need to be 'normalized' in such a way that each of them is of unit magnitude.
We have to find the appropriate $\gamma^0\equiv\gamma^t$. It needs to be checked that the $\gamma$ matrices satisfy Dirac algebra
$\{\gamma^{\mu},\gamma^{\nu}\}=2\delta^{\mu\nu}$.

Taking $t=1$, we obtain the following gamma matrices valid for a spin system in planar configuration on the square lattice.

\begin{equation}\label{Gx1}
\gamma^x=i\gamma^t\left( \begin{array}{cccc}
0&1&0&0\\
1&0&0&0\\
0&0&0&-1\\
0&0&-1&0
\end{array} \right)=
\left( \begin{array}{cccc}
0&i&0&0\\
-i&0&0&0\\
0&0&0&-i\\
0&0&i&0
\end{array} \right)
\end{equation}
\begin{equation}\label{Gy1}
\gamma^y=i\gamma^t\left( \begin{array}{cccc}
0&0&0&1\\
0&0&1&0\\
0&1&0&0\\
1&0&0&0
\end{array} \right)=
\left( \begin{array}{cccc}
0&0&0&i\\
0&0&-i&0\\
0&i&0&0\\
-i&0&0&0
\end{array} \right)
\end{equation}
with
\begin{equation}\label{Gt}
\gamma^t=\left( \begin{array}{cccc}
1&0&0&0\\
0&-1&0&0\\
0&0&1&0\\
0&0&0&-1
\end{array} \right)=(\gamma^t)^{-1}
\end{equation}

We aim this Dirac fermion system to produce a Chern-Simons term. For that we need the spectrum to be massive. 
We need to make sure that the Dirac masses from the first two sublattices and the second two sublattices have the same sign, otherwise the net Chern-Simons term vanishes.
Such mass term can be obtained in several different ways. First, using staggered sublattice chemical potential which represents some internal sublattice or microscopic degree of freedom which 
gives rise to effective chemical potential alternating in sign between the four sublattices of the square lattice. 
In this case the mass term from staggered chemical potential takes the form $H_M=\int d^2x \psi^{\dag}[\mathrm{diag}(m_s,-m_s,m_s,-m_s)]\psi$ where $[\mathrm{diag}(\cdots)]$ represents diagonal matrix
with the diagonal elements as given.

\begin{figure}
 \centering
 \includegraphics[scale=0.20]{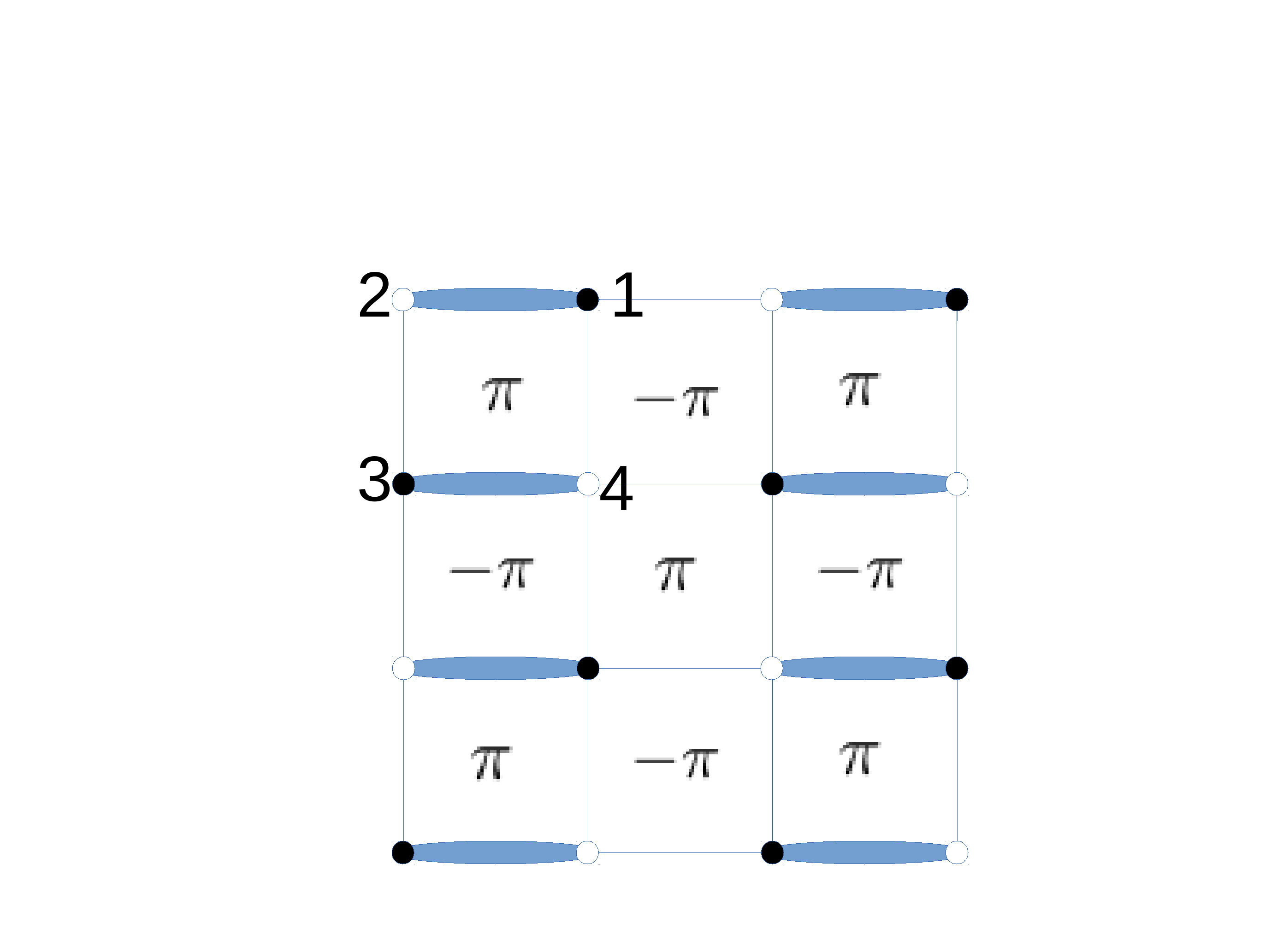}
  \caption{The columnar dimerized $\pi$ flux state of Affleck-Marston ansatz.}
\label{fig:dimerizedpifluxstate}
\end{figure}

The second method is to consider dimerization of the strength of the hopping integral $t$ in the tight-binding Hamiltonian Eq.(\ref{tightbindingH1}). Dimerization has been known to produce a mass for the Dirac fermions. We consider one of the simplest dimerization patterns as shown in the Fig. (\ref{fig:dimerizedpifluxstate}). 
The strength of the dimerized bonds is larger than the undimerized bonds. The link spin field and the flux per plaquatte remain the same as those of the original Affleck-Marston $\pi$-flux state.
This dimerization does open up gap on the Dirac points in the $4\times 4$ form of the Dirac theory which for the dimerization pattern shown in Fig. (\ref{fig:dimerizedpifluxstate}) takes the form

\begin{equation}\label{dimermass}
\mathcal{H}^{\mathrm{dimer}}_{M}=\gamma^t\left( \begin{array}{cccc}
0&m_0&0&0\\
m^*_0&0&0&0\\
0&0&0&m^s_0\\
0&0&{m^s_0}^*&0
\end{array} \right)
\end{equation}
where $m_0=-(\eta-1)\exp(-i\pi/4-ik_{Dx}),m^s_0=-(\eta-1)\exp(-i\pi/4+ik_{Dx})$ with $\eta>1$ is the relative strength of dimerized bond with respect to undimerized bond of unit strength.
Using $k_{Dx}=\pi/2$ from the location of the Dirac point, we obtain $m^s_0=-m_0=(\eta-1)\exp(-3i\pi/4)$. So, the Dirac mass Hamiltonian induced by dimerization Fig. (\ref{fig:dimerizedpifluxstate})
can be written as 

\begin{equation}
 H_M=\int d^2x\overline{\psi}\mathcal{H}^{\mathrm{dimer}}_{M}\psi=\int d^2x\overline{\psi}\gamma^t\frac{(\eta-1)}{\sqrt{2}}(\tau^x-\tau^y)\sigma^z\psi
\end{equation}
where $\tau$ is the Pauli matrix defined within each of the $1-2$ and $3-4$ sublattice pairs while $\sigma$ is the Pauli matrix connecting the two pairs.
In the rest of this derivation, we always imply

\begin{equation}
 H_M=\int d^2x\overline{\psi}\mathcal{H}_{M}\psi=\int d^2x\psi^{\dag}\mathbb{H}_{M} \psi
\end{equation}
where $\mathbb{H}_{M}=(\gamma^t)^{-1}\mathcal{H}_{M}$ is the Hermitian mass matrix derived directly from the tight-binding Hamiltonian.
 
Evaluating the mass Hamiltonian matrix Eq. (\ref{dimermass}) at the representative Dirac point, we obtain the following result for mass matrix.

\begin{widetext}
\begin{equation}\label{dimermass1}
\mathcal{H}^{\mathrm{dimer}}_{M}=\gamma^t\left( \begin{array}{cccc}
0&\frac{(\eta-1)}{\sqrt{2}}(1+i)&0&0\\
\frac{(\eta-1)}{\sqrt{2}}(1-i)&0&0&0\\
0&0&0&-\frac{(\eta-1)}{\sqrt{2}}(1+i)\\
0&0&-\frac{(\eta-1)}{\sqrt{2}}(1-i)&0
\end{array} \right)
\end{equation}
\end{widetext}
The above result however suggests that the masses are opposite in sign between the two (2-sublattice) pairs of the 4-sublattice theory (1-2 sublattice pair versus 3-4 sublattice pair). The net Chern-Simons term will be zero in this case.
To remedy this, we combine columnar or staggered dimerization with staggered sublattice potential where the staggerization takes 2-sublattice structure; the chemical potentials of sublattices 1 and 3 are equal
but opposite to the chemical potentials of sublattices 2 and 4. The resulting mass matrix is
\begin{widetext}
\begin{equation}\label{dimerstaggchempot}
\mathcal{H}^{\mathrm{dimer+stagg.chem.pot}}_{M}=\gamma^t\left( \begin{array}{cccc}
m_s&-\frac{(\eta-1)}{\sqrt{2}}(1+i)&0&0\\
-\frac{(\eta-1)}{\sqrt{2}}(1-i)&-m_s&0&0\\
0&0&m_s&\frac{(\eta-1)}{\sqrt{2}}(1+i)\\
0&0&\frac{(\eta-1)}{\sqrt{2}}(1-i)&-m_s
\end{array} \right)
\end{equation}
\end{widetext}
where $m_s$ is the mass from staggered sublattice chemical potential. It can readily be seen that the above mass matrix will give rise to nonzero net Chern-Simons term since the mass block-matrix of the 
$1-2$ sublattice pair is not opposite of that of the $3-4$ sublattice pair as the staggered sublattice potential-induced mass term matrix has precisely the same form and sign in both sectors.
If upon diagonalization the masses are of the same sign in both sector, which is indeed the case at least for certain range of parameters $m_s$, the the net Chern-Simons term is nonzero.
Performing this $4\times 4$ matrix diagonalization of the corresponding $\mathbb{H}_{M}$, we find that the Dirac mass Hamiltonian can be written as

\begin{equation}\label{Diracaction2x2columnardimer}
 H_m=\int d^2 x [\overline{\psi}'_{1}m_{\psi}\psi'_{1}+\overline{\psi}'_{2}m_{\psi}\psi'_{2}]
\end{equation}
where $\overline{\psi}'_{1(2)},\psi'_{1(2)}$ represent the first (second) pair of sublattice 1-sublattice-2 spinor $\psi'_1=(c'_1,c'_2)^T$ in diagonalized basis, with $\gamma^0=\tau^z$ and 
mass $m_{\psi}=\sqrt{m^2_{\mathrm{dimer}}+m^2_s}$ where $m_{\mathrm{dimer}}=\eta-1$ for both pairs, indicating that the two pairs have identical mass and thus nonzero net Chern-Simons term eventually.  
This result is nicely consistent with symmetry considerations because the combination of columnar or staggered dimerization plus staggered chemical potential completely breaks parity. 
It can also be checked that this dimerization plus staggered chemical potential breaks the invariance under time reversal times any lattice translation and this guarantees nonzero net
Chern-Simons term.


\begin{thebibliography}{1}
\bibitem{Caution}The result is obtained upon expansion in $\slashed{A}/(i\slashed{\partial}+m)$ at order $\mathcal{O}(\slashed{A}^2)$.
\end{thebibliography}
\end{document}